# External pressure dependence of Granular Orifice Flow: transition to Beverloo flow

Zheng Peng(彭政)[1]*, Jiangmeng Zhou(周江盟)[1], Jiahao Zhou(周家豪)[1], Yuan Miao(缪源)[1], Liyu Cheng(程礼渝)[1], Yimin Jiang(蒋亦民)[1], Meiying Hou(厚美瑛)[2]†

[1]*School of Physics and Electronics, Central South University, Changsha 410012, China*

[2]*Key Laboratory of Soft Matter Physics, Beijing National Laboratory for Condense Matter Physics, Institute of Physics,Chinese Academy of Sciences, Beijing 100190, China*

Abstract

In this paper, we have designed and employed a suspended-wall silo to remove Janssen effect in order to explore directly the local pressure dependence of Granular Orifice Flow (GOF) systematically. We find that as Janssen effect is removed, the flow rate $Q$ changes linearly with the external pressure. The slope $\alpha$ of the linear change decays exponentially with the ratio of the silo size and the size of the orifice $\Phi/D$, which suggests the existence of a characteristic ratio $\lambda$ (~2.4). When $\Phi/D > \lambda$, $\alpha$ gradually decays to zero, and the effect of external pressure on the GOF becomes negligible, where the Beverloo law retrieves. Our results show that Janssen effect is not a determining factor of the constant rate of GOF, although it may contribute to shield the top load. The key parameter in GOF is $\Phi/D$. In small $\Phi/D$, the flow rate of GOF can be directly adjusted by the external pressure via our suspended-wall setup, which may be useful to the transportation of granules in microgravity environment where the gravity-driven Beverloo law is disabled.

* zpeng@csu.edu.cn
† mayhou@iphy.ac.cn

## I. INTRODUCTION

With important engineering applications, such as storage and discharge of solids in a silo, transportation of raw materials in a pipe or conveyor belt, Granular Orifice Flow (GOF) has been extensively studied for decades [1–9]. A well-known phenomenon of GOF is that its flow rate is a time-independent constant, and has been used as a timer with accuracy better than that of the clepsydra [10, 11]. The Beverloo empirical formula of a three-dimensional GOF given in 1961[11, 12], for a silo with a bottom orifice of size $D$, provided silo height $H > 2D$ and silo diameter $\Phi > 2D$, is

$$Q = C\rho\sqrt{g}(D - D_B)^{\frac{5}{2}} \qquad (1)$$

where $\rho$ is the bulk density of the granular material, $g$ is the gravitational acceleration, and $C$ and $D_B$ are empirical constants. In this empirical formula, the flow rate $Q$, driven by gravity, is only a function of the orifice size $D$, and is independent of the remaining filling height of granules in the silo.

The most intuitive understanding of the independence of the flow rate on silo height is based on the so-called Janssen effect [13, 14]: the weight of the granules and the loads on the top is partially shielded by the side wall friction, which leads the pressure at the bottom to a saturation constant, i.e., bottom pressure becomes height independent. Therefore, the GOF flux is independent of the filling height [15]. The concept of the Janssen effect is proposed to describe the stress distribution of a static silo. Whether the concept is still applicable to dynamic processes is, however, questionable when the silo is discharged through the bottom orifice [16]. Moreover, the interpretation based on the Janssen effect implies that bottom pressure is a determining factor of GOF, which is inconsistent with some recent experimental results [16–18]. Aguirre et al. performed a two-dimensional GOF experiment based on the conveyor belt and found that the flow rate and bottom pressure were independent [16, 17]. Perge et al. measured the distribution of the bottom pressure by gravity-driven three-dimensional GOF experiments. It was experimentally verified that the dynamic Janssen effect was present during the silo discharging process, and the pressure changed near the orifice region did not change the flow rate, that is, the pressure and the flow rate were not one-to-one correspondent [18]. Free Fall Arch (FFA) is assumed to explain the independence of flow rate of GOF from pressure [11, 19–21]. It assumes that an arch is formed in the vicinity of the orifice. The arch withstands all the pressure above it, thus ensuring that under the arch, the granules free fall under gravity, that is, the stress distribution in the silo become discontinuous above and under the arch. For an orifice of diameter $D$, the typical velocity of the granules falling from the arch through the orifice is $v \propto \sqrt{gD}$, and the cross-section area of the orifice is $s \propto D^2$, therefore, the flow rate of three-dimensional GOF is $Q \propto vs \propto D^{\frac{5}{2}}$, which is independent of the pressure above the arch. This assumption sounds reasonable and even explains the 5/2 power relationship between $Q$ and exit size $D$ in the Beverloo formula. However, whether FFA exists in GOF is still controversial [21, 22]. Recent simulations show that FFA, even if it exists, may not be a typical arch with discontinuous stress field [21].

Regardless of the existence of FFA, the flow rate of GOF is not controlled by pressure, or says the robustness of Beverloo's formula is verified in many experiments. Generally speaking, the flow rate of any flowable material through an orifice should be controlled by the pressure near the orifice. The fact that the flow rate of GOF is independent of the local pressure is mysterious. Is the flow rate of GOF really independent of bottom pressure in all cases? Madrid et al. [23] applied an additional load on the top of a silo and found that the rate was affected by the top load. However, the phenomena were significant only when $H/\Phi < 6$; at $H/\Phi > 6$, different loads showed the same discharge rate (see fig.2 in ref [23]). We deem that the finding in ref [23] confirms the existence of dynamic Janssen effect; at $H/\Phi > 6$, the top loads were shielded by the dynamic Janssen effect and the local pressure near the orifice tended to be the same which results in the same flow rate. In other words, in a typical silo, due to the dynamic Janssen effect, it's hard to adjust the local pressure near the orifice even by loading on top, and therefore hard to explore the pressure dependence of GOF systematically. In this paper, we try to answer the question: is it possible to remove the shielding of the dynamic Janssen effect and study the pressure effects of GOF over a wider range of pressure?

In our previous work [24], we propose a suspended-wall experimental setup that can effectively remove the Janssen effect and load the top pressure fully to the bottom plate to study the pressure effect of the GOF. In this paper, based on a similar experimental framework as that in ref [24], we systematically explored the influence of external load on the flow rate $Q$ of GOF for different silo sizes $\Phi$ and different orifice sizes $D$ to test the robustness of Beverloo formula. It is found that $Q$ increases linearly with the pressure within the experimentally range for all cases with various $\Phi$ and $D$, and all the slopes could collapse well into a curve if scaled by $\Phi/D$. Moreover, the slope α as a function of $\Phi/D$ follows an exponential decay with a characteristic ratio λ (~ 2.4). When $\Phi/D < \lambda$, the GOF is obviously affected by the pressure, with the flow rate $Q$ increases linearly with the pressure. While at $\Phi/D \gg \lambda$, the influence of pressure is negligible and the Beverloo formula is revived in gravitational field, i.e., this external loaded GOF is transitioned to Beverloo flow.

**II. EXPERIMENTAL SETUP**

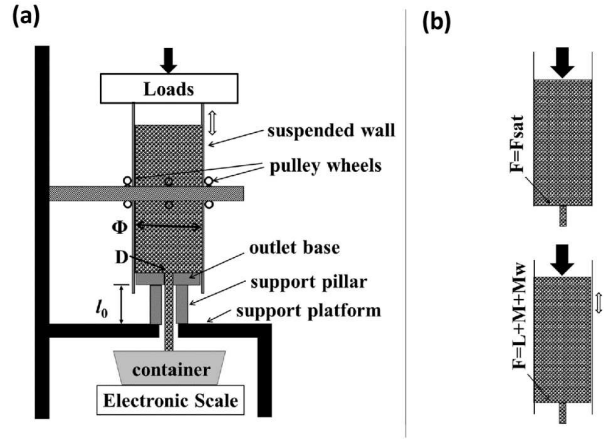

FIG. 1: (a): Sketch of the experimental setup. (b): schematic diagram of the bottom pressing force of a typical silo(above) and a suspended-wall silo(below).

As shown in figure 1 (a), the experimental setup is a vertical silo which mainly composed of two mechanically independent parts: a moveable sidewall and an outlet base. The sidewall is an aluminum hollow cylinder, which horizontally restricted by pulley wheels such that it can be moved freely in the vertical direction, with inner diameters $\Phi$ (= 44, 56, 80, 150 mm); while the base is a thin aluminum plate with its radius 1mm smaller than that of sidewall and an orifice of diameter $D$ (= 14, 16, 18, 20, 22 mm, for each $\Phi$) in the center. Since the granular materials used in the experiment are nearly mono-disperse glass beads (density 2.46$g/cm^3$) with diameter $d \sim 3.0$ mm, the gap between the base and sidewall is smaller than half of the bead size, which ensuring no beads would stuck in the gap during the experiment process. A rigid hollow pillar with length $l_0$ = 120 mm, connecting the base and a rigid platform, supports the weight of whole silo including load on the top, and provide a free distance $l_0$ for sidewall to move. Before initiating the experiment, we have the outlet plugged, fill the silo to an initial height $H_i$ (∼ 60 cm), and pull the sidewall to an initial position to ensure the bottom of sidewall is aligning with that of the base. Since the sidewall and base of the silo are mechanically separated, the sidewall will be suspended due to the static friction with the granules. Different from the typical Janssen effect that the sidewall sustains partial weight of granules in the silo, in this design the weight of the sidewall including the load on the top is sustained by the granules and eventually balanced by the support of silo base. Hence, by allowing the sidewall to move freely and vertically, the screening of sidewall in the usual silo, namely Janssen effect is removed and the total pressing force on the base $F$ is then a sum of the load $L$, the weight of granules $M$ and weight of the wall $M_W$ – a quantity that can be experimentally controlled and varied. Figure 1 (b) is a schematic diagram of the bottom pressing force of a typical silo and a suspended-wall silo. For a typical silo shown in the figure above, after being filled to a certain height, the force at the bottom will saturate due to the Janssen effect: $F = F_{sat}$; while for the suspended-wall silo shown in the figure below, the bottom force $F = L+M+M_W$, which will never saturate. We can increase $F$ by increasing $L$, $M$ or $M_W$. However, increasing $M$ is hard to operate in experiment. For instance, for a silo of $\Phi = 150$ mm, a weight of $M$=80kg would correspond to a filling height over 3 meters, while the same $F$ can be realized easily by adding a load $L$=80kg on the silo top. In this work, we choose varying $L$ and keep the $M$ and $M_w$ fixed.

In the previous study [25], the maximum static fiction force of a suspended sidewall with the granules was measured by vertical pulling and pushing the wall of a silo. It was found that the pushing force was one order of magnitude larger than the pulling force, and increased exponentially with the filling mass (or height). It should be noted that the difference between the pulling force and the pushing force is caused by the reverse of friction force on the sidewall. Recently, a similar phenomenon was explored as the reverse Janssen effect [26]. In our current experiment, with a suitable filling height the sidewall can remain stationary with the granules even when the load on the top of the sidewall exceeds 80 kg (the maximum load in the experiment). From the point of view of force balance, whether the load is applied to the top of the sidewall or to the top of the granules does not make a difference to the total pressure at the bottom of the silo. In fact, we have tried both loading methods in our experiments and found no observable effects on the experimental results (flow rate). For the convenience of experimental operation, the former loading method was employed in the following experiments.

When the outlet is opened, the beads fall in a container and the mass $m(t)$ of beads in the container at time $t$ is measured by an electronic scale of the precision ±1 g recording at 10 Hz. Except for a short initial unstable flow, the slope of $m(t)$ is used to calculate the flow rate $Q$.

### III. EXPERIMENTAL RESULT AND DISCUSSION

As the silo is drained, the suspended sidewall descends synchronously. It should be noted that if the orifice is suddenly plugged during drainage, the descending of the sidewall will stop immediately, which confirms that there is no relative sliding between the sidewall and the granules during the descent process. After descending a free distance $l_0$ the sidewall landed on the rigid support platform, and then the forced GOF was turned into a usual Beverloo flow. Figure 2 is a typical curve of mass $m(t)$ increasing with time $t$ in the suspended wall experiment. The flat portions of the curve correspond to the states before the flow started and after the flow ended, respectively. In the flowing regime, the curve is clearly divided into two straight lines, and the inflection point corresponds to the moment when the sidewall lands at the bottom plate. The slopes of the two straight lines correspond to the flow rates of the two GOFs, that is, the $Q_F$ of the forced flow in the suspended wall condition and the $Q_B$ of the Beverloo flow in the fixed wall condition. From the difference in the slopes, $Q_F$ is indeed significantly larger than $Q_B$.

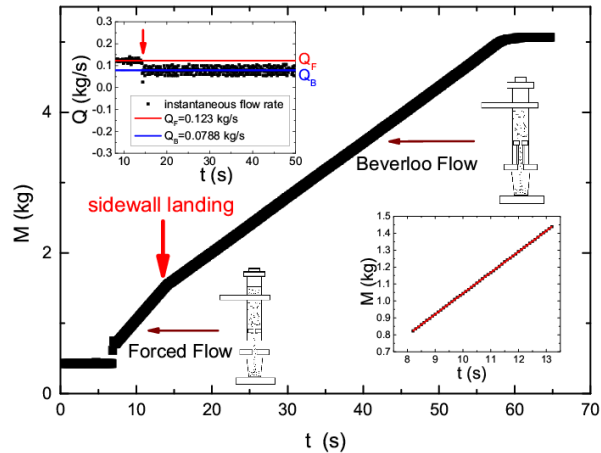

FIG. 2: A typical curve of mass $m(t)$ increasing with time $t$ in the suspended wall experiment. The curve is divided into two regimes – forced flow regime and Beverloo flow regime, by "sidewall landing". Inset (lower right): a partial enlarged view of the forced flow regime in the main diagram, the solid line is a linear fitting; inset (upper left): instantaneous flow rate calculated by point-by-point differential, the two solid lines are the slopes of the two straight lines in the main graph, which corresponding to the two flow rates $Q_F$ and $Q_B$ respectively.

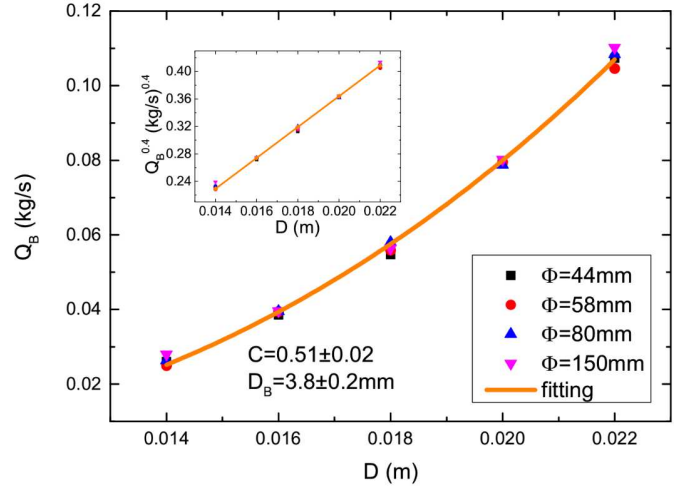

FIG. 3: In fixed-wall regime (Beverloo flow), flow rate $Q_B$ versus orifice diameters $D$ for different $\Phi$. Inset is $Q_B^{0.4}$ versus $D$. Solid line is fitting with equation (1).

The lower right inset of Fig. 2 is a partial enlarged view of the $m(t)$ curve of the forced flow regime in the main diagram. The good linearity of the curve indicates that $Q_F$ is a constant throughout the regime. In fact, during forced flow, the mass $M$ in the silo and the total bottom pressing force $F$ is decreasing due to the outflow of granules (the maximum reduction of $M$ is $\frac{\rho \pi D^2 l_0}{4}$ in the forced flow regime). However, since the main contribution to $F$ is $M_W$, $L$ and the remaining $M$ in the silo, the slight change of $F$ during the forced flow has no significant effect on the flow rate $Q_F$ as shown in the inset. Hence we use the initial $F$ as the pressing force corresponding to $Q_F$. The upper left inset of Fig. 2 is the instantaneous flow rate over time calculated from the data of main graph by point-by-point differential. Although the instantaneous flow rates fluctuate due to the numerical differentiation, two significantly different stable flow rates $Q_F$ and $Q_B$ are observed. The two solid lines in the inset are the slopes of the two straight lines in the main graph, which are consistent with the instantaneous flow rates.

As the sidewall landed on the rigid support platform, the suspended-wall GOF turned to a fixed-wall GOF, i.e. Beverloo Flow, as can be seen in Figure 3. The flow rate $Q_B$ vs. $D$ under different $\Phi$ is plotted in Figure 3. Shown in the figure, $Q_B$ is only a function of $D$ and is independent of $\Phi$. The upper left inset is the relationship between $Q_B^{0.4}$ and $D$. A good linear fit in the inset shows that the relationship between $Q_B$ and $D$ does follow the Beverloo empirical formula. The solid line in the figure is the fitting with equation (1), and the fitting parameters $D_B = 3.8 \pm 0.2$mm, $C = 0.51 \pm 0.02$, which are in line with the empirical value range of Beverloo Flow [11].

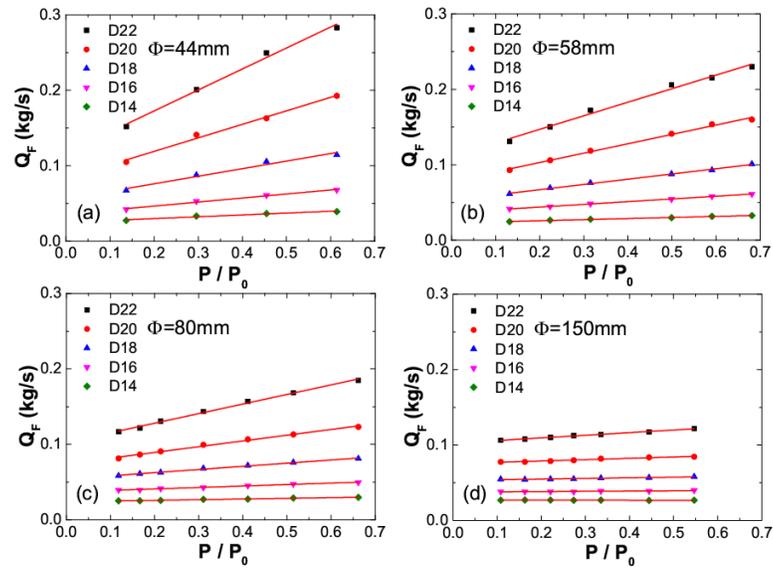

FIG. 4: In suspended-wall regime(Forced Flow), the relationship between $Q_F$ and $P/P_0$ at different orifice $D$. (a)-(d) correspond to $\Phi$ = 44, 58, 80, 150 mm, respectively. The solid lines are linear fittings.

In the Forced Flow (FF) regime, the flow rate $Q_F$ is not only a function of $D$, but also affected by the diameter of the silo $\Phi$ and the applied pressure $P$. Figure 4 show the relationship between $Q_F$ and $P/P_0$ at different orifice $D$, where $P_0$ is atmosphere pressure, and (a)-(d) correspond to the cases of $\Phi$ = 44, 58, 80, 150 mm, respectively. It's shown that for all $\Phi$ and $D$, the flow rates $Q_F$ of FF increase linearly with $P$, and can be expressed as:

$$Q_F = Q_0 \left(1 + \alpha \frac{P}{P_0}\right) \qquad (2)$$

where $\alpha$ and $Q_0$ are fitting parameters.

Figure 5 shows the variation of the fitting parameters $Q_0$ with $D$ at different $\Phi$. Since $Q_0$ is the intercept of the linear fitting of figure 4, it has a certain fitting error. However, within the range of fitting error, $Q_0$ is only a function of $D$ and is independent of $\Phi$, which is consistent with the variation relationship of $Q_B$ and $D$ in the case of fixed wall shown in fig.3. That is, in the gravitational field, $Q_0$ tends to Beverloo's empirical formula, i.e.

$$Q_0 = C_0 \rho \sqrt{g}(D - D_B)^{\frac{5}{2}} \qquad (3)$$

which is also confirmed by the linear relation between $Q_0^{0.4}$ and $D$ shown in the inset. It should be noted that $Q_0$ is entirely the result of linear fits from the data of fig.4, therefore its relationship to $D$ is not a priori.

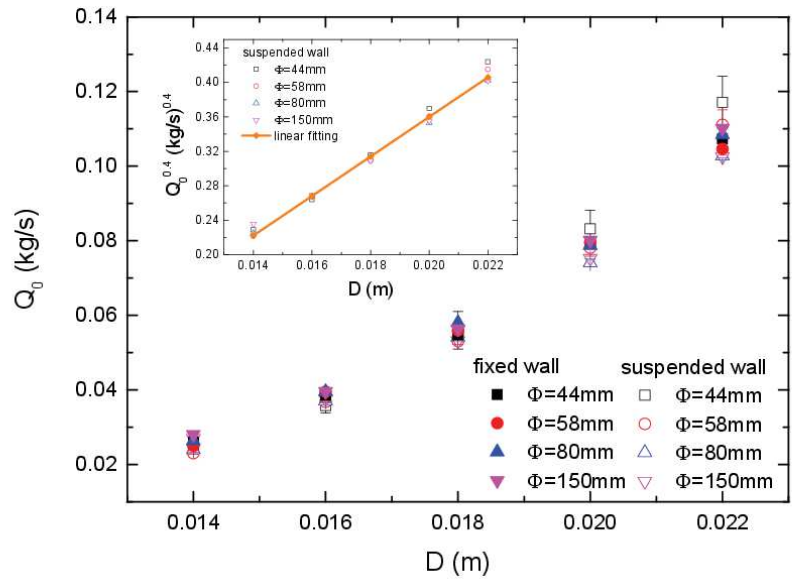

FIG. 5: The intercept of linear fitting of data in fig.4, $Q_0$ varies with $D$ at different $\Phi$. the $Q_B$ of fixed-wall flow (solid symbols) from fig.3 are added in the graph for comparison. Inset is another drawing with $Q_0^{0.4}$ versus $D$, the solid line is a linear fitting.

It can be seen from fig.4 that for a fixed $\Phi$, if $D$ is larger, the slope $\alpha$ of the linear fitting is larger;

and when $\Phi$ is increased, the slope $\alpha$ starts to become smaller for the same size $D$. This suggests that there is a competitive relationship between the influence of $\Phi$ and $D$ on $\alpha$. Figure 6 shows the relationship between α and $\Phi/D$. As shown in the figure, all of the $\alpha$'s are well overlapped and exhibit an exponential decay with increasing $\Phi/D$. For a sufficiently large $\Phi/D$, $\alpha$ tends to zero. As shown in the single logarithmic coordinates of the inset, ln$\alpha$ exhibits a linear relationship with $\Phi/D$, indicating the presence of a characteristic ratio λ. The experimental results of $\alpha$ can be well fitted by the formula $\alpha = Ae^{-\frac{\Phi/D}{\lambda}}$, with fitting parameter $A = 4.8 \pm 0.4$ and the characteristic ratio $\lambda = 2.4 \pm 0.1$.

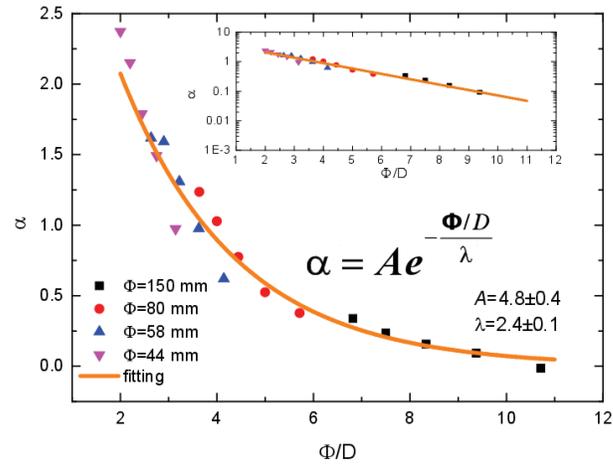

FIG. 6: The slope of linear fitting of data in fig.4, $\alpha$ varies with $\Phi/D$. Inset is another drawing in single logarithmic coordinates. Solid line is fitting with formula $\alpha = Ae^{-\frac{\Phi/D}{\lambda}}$, with fitting parameter A= 4.8 ± 0.4 and λ = 2.4 ± 0.1.

Usually influence of grain diameter $d$ on the flow rate is included by writing the constant $D_B=kd$ of Eqs.(1,3), with the concept of "empty annulus", see ref[11]. Such a correction could be also present in the exponential model for $\alpha$, namely replacing the orifice diameter $D$ by $D-k'd$ in the fig.6. with $k'$ a fitting parameter. As it leads only detailed corrections and doesn't alter our main conclusions, the effect of grain sizes is not considered in this work. More systematic measurements for various $d$ are required for studying it.

In Fig.5, the curves of suspended wall for small $\Phi$'s (especially for Φ=44mm) have a tendency to deviate from the collapsed curve. For all data of suspended wall except for that of Φ=44mm in

fig.5, the best fitting of Eq. (3) gives $D_B$ = 4.3mm. Therefore, we can calculate $C_0$ directly for each $\Phi$ and D via $C_0 = Q_0/[\rho\sqrt{g}(D - D_B)^{\frac{5}{2}}]$ according to Eq. (3), to demonstrate the deviation. By taking $\rho$ = 1500kg/m$^3$ and $D_B$ = 4.3mm, $C_0$ as functions of $\Phi/D$ is shown in figure 7. As shown in the figure, the parameter $C_0$ is almost a constant 0.53 (consistent with the fitting value), and showing a tendency to increase when the ratio $\Phi/D < \lambda$, which could be explained by an onset to the vertical pipe flow.

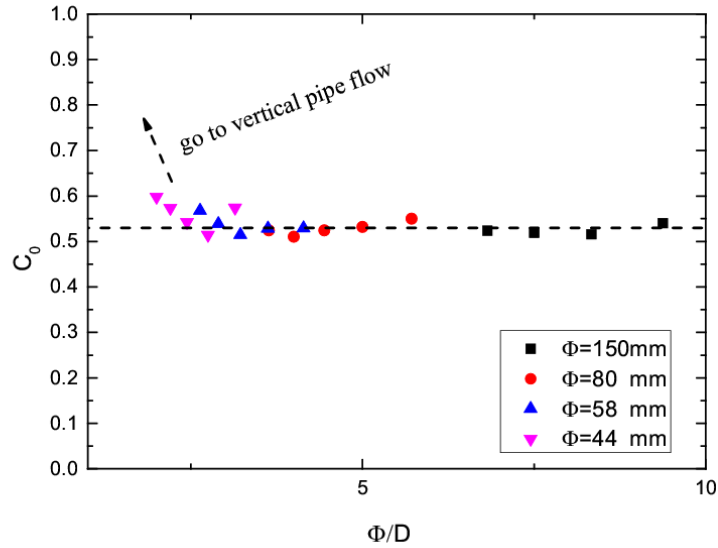

FIG. 7: Values of the parameter $C_0$ calculated from Eq. (3), for various $\Phi$, $D$.

The presence of the characteristic ratio $\lambda$ means that when $\Phi/D < \lambda$, the GOF is significantly affected by the pressure, and the flow rate increases linearly with the pressure. In fact, the lower limit of $\Phi/D$ is 1, at which limit GOF turns to pipe flow, which is a flow state that is obviously affected by pressure. When $\Phi/D > \lambda$, $\alpha$ gradually decays to zero, and the effect of pressure on the GOF becomes negligible. At this limit, the flow rate $Q$ shown in the equation (2) turns to $Q_0$, a function of $D$ (shown in equation (3)), and the GOF turns to Beverloo flow. In other words, we have found a characteristic ratio $\lambda$ of the GOF to describe the transition from forced flow to Beverloo Flow. For GOF, when $\Phi/D$ is close to $\lambda$, the flow rate is the pressure-controlled Forced Flow; and when $\Phi/D$ is large enough, the flow rate turns to pressure-independent Beverloo Flow. If $\Phi/D$ is large enough, even when the Jansen effect is completely eliminated, one can increase the average pressure at the bottom at will, and the flow of GOF will maintain Beverloo flow unchanged, which shows the robustness of Beverloo Flow.

We assume simply in this work the case that tangential force along whole silo wall is upwards, of which summation equals to the wall weight. It is possible that the force changes sign, for example, directing downwards in an upper part, upwards in remaining lower part. Clearly this is a complicated mechanical status with a Janssen effect in upper and a reverse one in lower, and requires further experiments to study forces along the wall, which beyond the scope of this work. However, as it doesn't influence the bottom pressure, our results will remain valid even if the complication exists.

The observed exponential transition to the robust Beverloo flow for silo with vertically suspended wall can be reproduced by the DEM simulation works [27, 28]. A remarkable result of the DEM works is that rolling friction among grains is found crucial for simulating the observed variations of flow rate with top loads shown by the Fig.4 (neglecting the rolling friction the variations are heavily overestimated). Notably for DEM simulations of silo discharges, the Janssen effect could be removed also by taking friction between sidewall and granulates equal to zero, and a same exponential transition as shown by the Fig.6 can be also reproduced [28]. Why the rolling friction is important for DEM simulating the present measurements is an interesting topic. It is worth to note that in the DEM works [29-31], the importance of rolling friction for reproducing another basic granular property called "rate independence" in soil mechanics is claimed. The rate independence means that stress-strain behaviors are independent on strain rate, and the robustness of Beverloo flow means that stress strength and flow rate are uncoupled. That both experiment facts are related to the rolling friction is considered reasonable.

It is worth pointing out that the existence of the characteristic parameter in the transition of granular medium is not rare. For example, when an intruder approaching the solid bottom boundary of a granular medium, the penetration resistance near the boundary is found increasing exponentially with decreasing distance, which may be related to the characteristic length of "jamming" caused by the locally applied stress [32]. When studying the criticality of the dilute-to-dense transition in granular flow, it's found that the probability function for the flow remaining dilute decays exponentially with the waiting time [33]. The transition follows exponential decay and has a characteristic parameter that appears to be a property of granular media. However, which features of the granular media are related to the characteristic parameters and whether there is a relationship between the characteristic parameters is worthy further research.

## IV. CONCLUSION

In this paper, by using a silo with a suspended wall, we removed the Janssen effect and effectively loaded the top pressure to the orifice at the silo bottom, explored the pressure effect of GOF at various silo sizes $\Phi$ and orifice sizes $D$. It is found that for a silo with given $\Phi$, flow rate $Q$ increase linearly with increasing pressure $P$, and the slope $\alpha$ of the linear increase is larger for a larger $D$. This indicates that increasing the pressure at the orifice does increase the flow rate effectively, i.e., the flow rate of the GOF can be controlled by pressure. As $\Phi$ increases, the slope $\alpha$ gradually decrease at same $D$, which indicates that increasing $\Phi$ can effectively suppress the pressure effect of GOF. Through analysis, it is found that for various sizes of $\Phi$ and $D$, the relationship between $\alpha$ and $\Phi/D$ can be well superposed on a curve and follows the exponential decay law. This indicates the presence of a characteristic ratio $\lambda$ (~2.4) in our granular system, which describes the GOF transition from forced flow to Beverloo Flow.

We are entering the space age, how to transport granular matter in the microgravity environment, such as how to make them effectively flow out of an orifice in space will be a research topic we have to face. In the microgravity environment, the gravity-driven Beverloo formula is obviously no longer applicable. In this case, the particles should pass through the orifice by applying external pressure. Therefore, effectively applying pressure to the vicinity of the orifice and exploring the pressure-driven GOF law (e.g., with our suspended wall equipment) will have more important research significance.

## ACKNOWLEDGMENTS

This work was partially supported by the National Natural Science Foundation of China (Grant No. U1738120) and Central South University Education and Teaching Reform Research Project (Grant No. 2020jy076).

## DATA AVAILABILITY

The data that support the findings of this study are available from the corresponding author upon reasonable request.

[1] C. Mankoc, A. Janda, R. Arvalo, M. Pastor, I. Zuriguel, A. Garcimartin, and D. Maza, Granular Matter 9, 407 (2007).

[2] D. Hernández-Delfin, T. Pongó, K. To, T. Börzsönyi, and R. C. Hidalgo, Phys. Rev. E 102, 042902 (2020).

[3] A. Janda, I. Zuriguel, and D. Maza, Phys. Rev. Lett. 108, 248001 (2012).

[4] J. R. Darias, M. A. Madrid, and L. A. Pugnaloni, Phys. Rev. E 101, 052905 (2020).

[5] A. Nicolas, A. Garcimartn, and I. Zuriguel, Phys. Rev. Lett. 120, 198002 (2018).

[6] C. C. Thomas and D. J. Durian, Phys. Rev. Lett. 114, 178001 (2015).

[7] K. To, Y. Yen, Y.-K. Mo, and J.-R. Huang, Phys. Rev. E 100, 012906 (2019).

[8] D. Gella, I. Zuriguel, and D. Maza, Phys. Rev. Lett. 121, 138001(2018).

[9] A. M. Cervantes-Álvarez, S. Hidalgo-Caballero and F. Pacheco-Vázquez, Phys. Fluids 30, 043302 (2018).

[10] L. Staron, P.-Y. Lagre, and S. Popinet, Phys. Fluids 24, 103301 (2012).

[11] R. M. Nedderman, Statistics and Kinematics of Granular Materials (Cambridge University Press, Cambridge, England, 1992). (chapter 10)

[12] W. A. Beverloo, H. A. Leniger, and J. J. Van de Velde, Chem. Eng. Sci. 15, 260 (1961).

[13] H. A. Janssen, Zeitschr. d. Vereines deutscher Ingenieure 39,1045 (1895).

[14] J. Duran, Sands, Powders and Grains: An Introduction to the Physics of Granular Materials (Springer-Verlag, Berlin, 2000).

[15] H. M. Jaeger, S. R. Nagel, and R. P. Behringer, Rev. Mod.Phys. 68, 1259 (1996)

[16] M. A. Aguirre, J. G. Grande, A. Calvo, L. A. Pugnaloni, and J.-C. Geminard, Phys. Rev. E 83, 061305 (2011).

[17] M. A. Aguirre, J. G. Grande, A. Calvo, L. A. Pugnaloni, and J.-C. Geminard, Phys. Rev. Lett. 104, 238002 (2010).

[18] C. Perge, M. A. Aguirre, P. A. Gago, L. A. Pugnaloni, D. Le Tourneau, and J. C. Geminard, Phys. Rev. E 85, 021303 (2012).

[19] Y. Tian, P. Lin, S. Zhang, C.L. Wang, J.F. Wan, L. Yang, Advanced Powder Technology 26, 1191(2015).

[20] P. Lin, S. Zhang, J. Qi, Y.M. Xing, L. Yang, Physica A 417, 29(2015)

[21] S. M. Rubio-Largo, A. Janda, D. Maza, I. Zuriguel, and R. C. Hidalgo, Phys. Rev. Lett. 114, 238002 (2015).

[22] D. J. Van Zuilichem, N. D. Van Egmond, and J. G. DeSwart, Powder Technol. 10, 161 (1974).

[23] M. A. Madrid, J. R. Darias and L. A. Pugnaloni, EPL 123, 14004 (2018)

[24] Z. Peng, H Zheng and Y. Jiang, arXiv:0908.0258v3, https://arxiv.org/abs/0908.0258 (2009).

[25] Z. Peng, X. Li, L. Jiang, L.Fu, and Y. Jiang, Acta Phys. Sin. 58(3), 2090 (2009).

ACCEPTED MANUSCRIPT

Physics of Fluids

This is the author's peer reviewed, accepted manuscript. However, the online version of record will be different from this version once it has been copyedited and typeset.
PLEASE CITE THIS ARTICLE AS DOI: 10.1063/5.0048357
[26] S. Mahajan, M. Tennenbaum, S. N. Pathak, D. Baxter, X. Fan, P. Padilla, C. Anderson, A. Fernandez-Nieves and M. Pica Ciamarra Phys. Rev. Lett. 124, 128002 (2020)

[27] Shunying Ji, Siqiang Wang, Zheng Peng. Powder Technology 356, 702 (2019).

[28] Sheng Zhang, private communication.

[29] Chuang Zhao, Chengbo Li, Physica A 460, 44 (2016).

[30] Chuang Zhao, Chengbo Li, Lin Hu, Physica A 492, 181 (2018).

[31] Chuang Zhao, Yinghao Luo, Lin Hu, Chengbo Li, Granular Matter 20, 66 (2018).

[32] M. B. Stone, D. P. Bernstein, R. Barry, M. D. Pelc, Y. Tsui and P. Schiffer, Nature 427, 503(2004).

[33] J. Zhong, M. Hou, Q. Shi and K. Lu, J. Phys.: Condens. Matter 18, 2789 (2006).

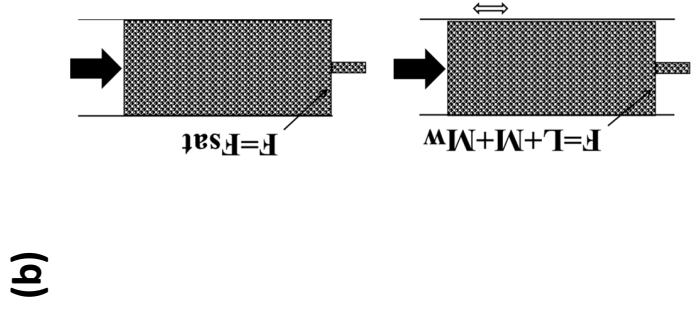

$F=F_{sat}$

$F=L+M+M_w$

(b)

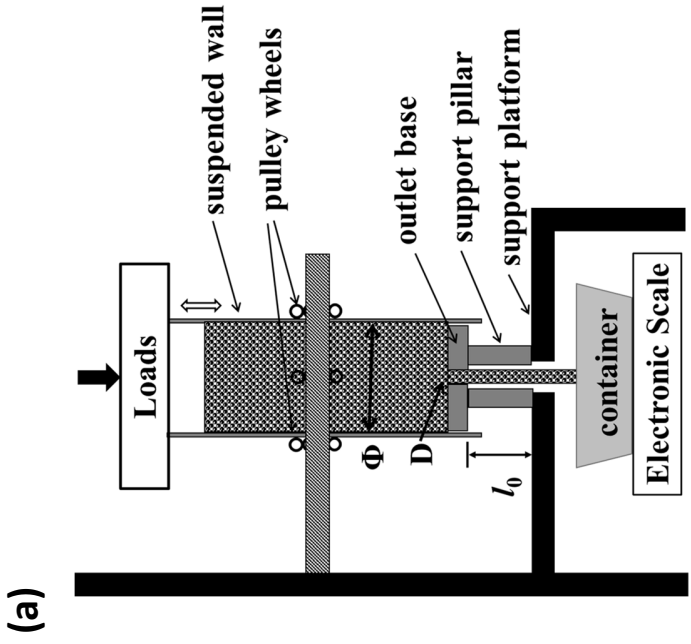

Loads — suspended wall — pulley wheels — outlet base — support pillar — support platform — container — Electronic Scale — $\Phi$ — D — $l_0$

(a)

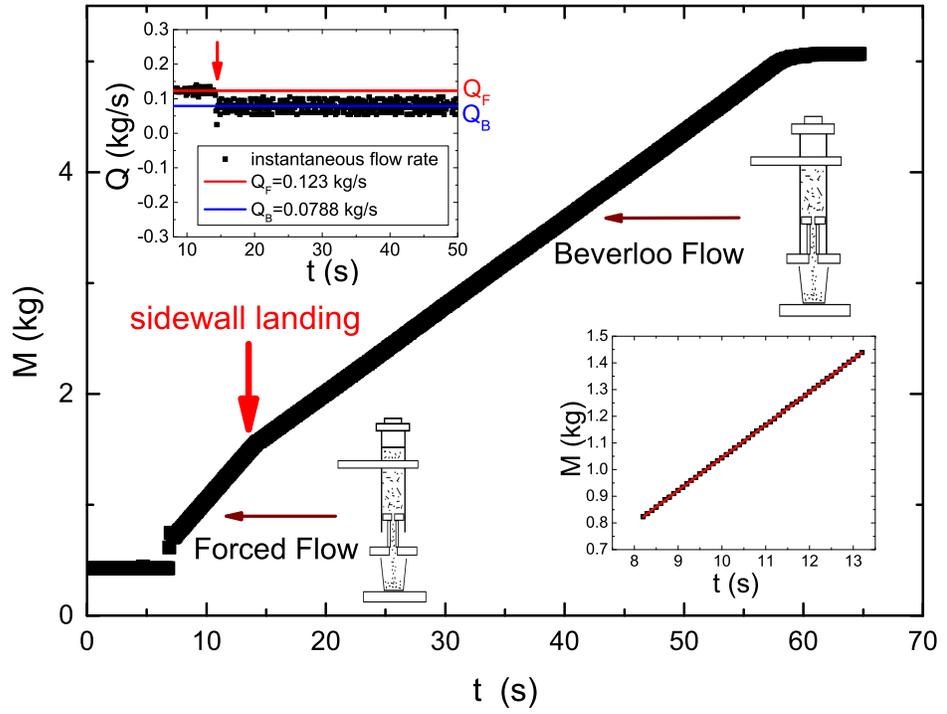

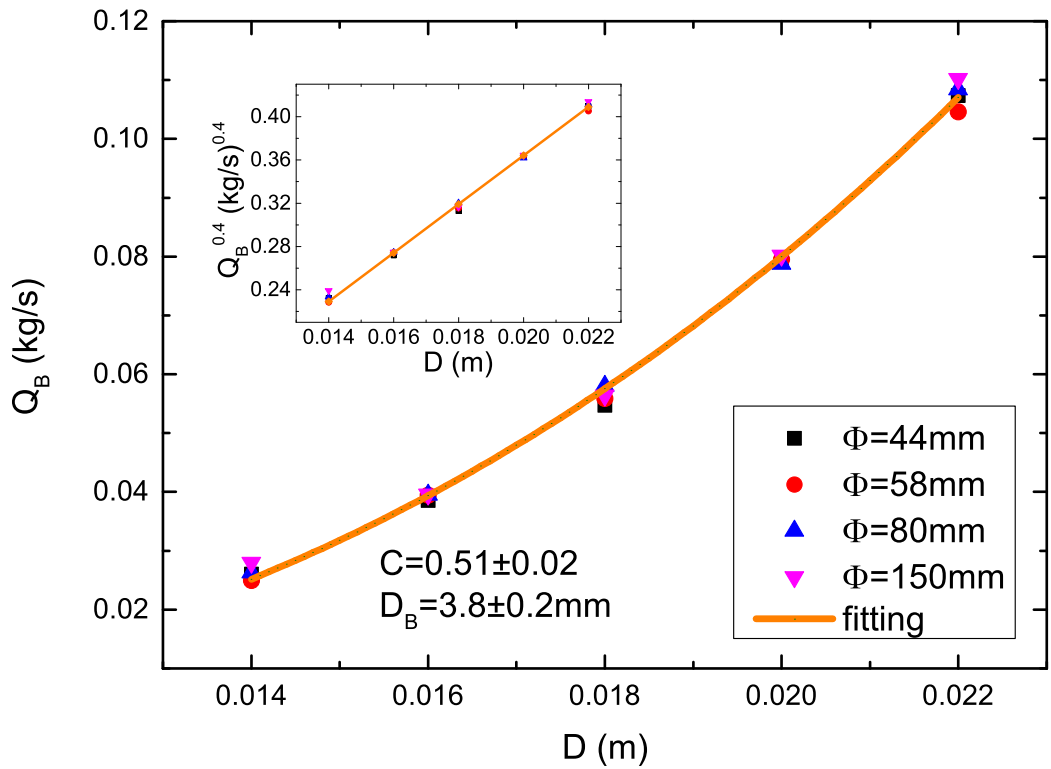

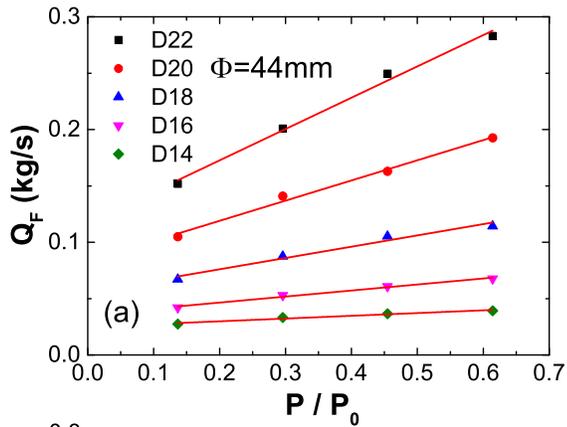
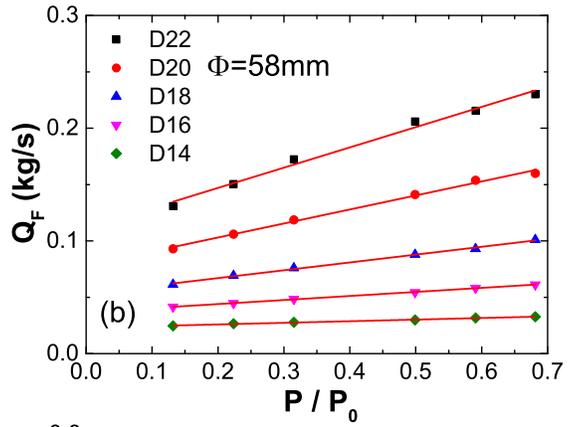
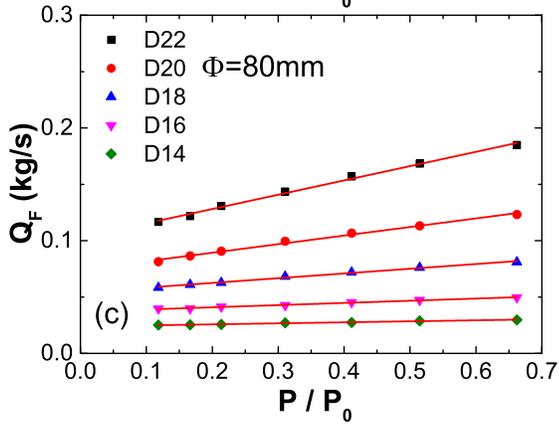
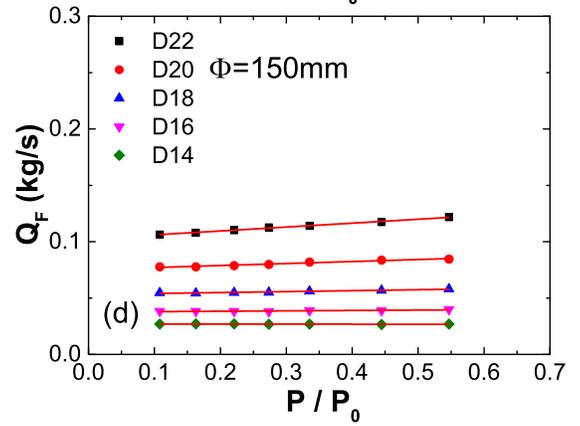

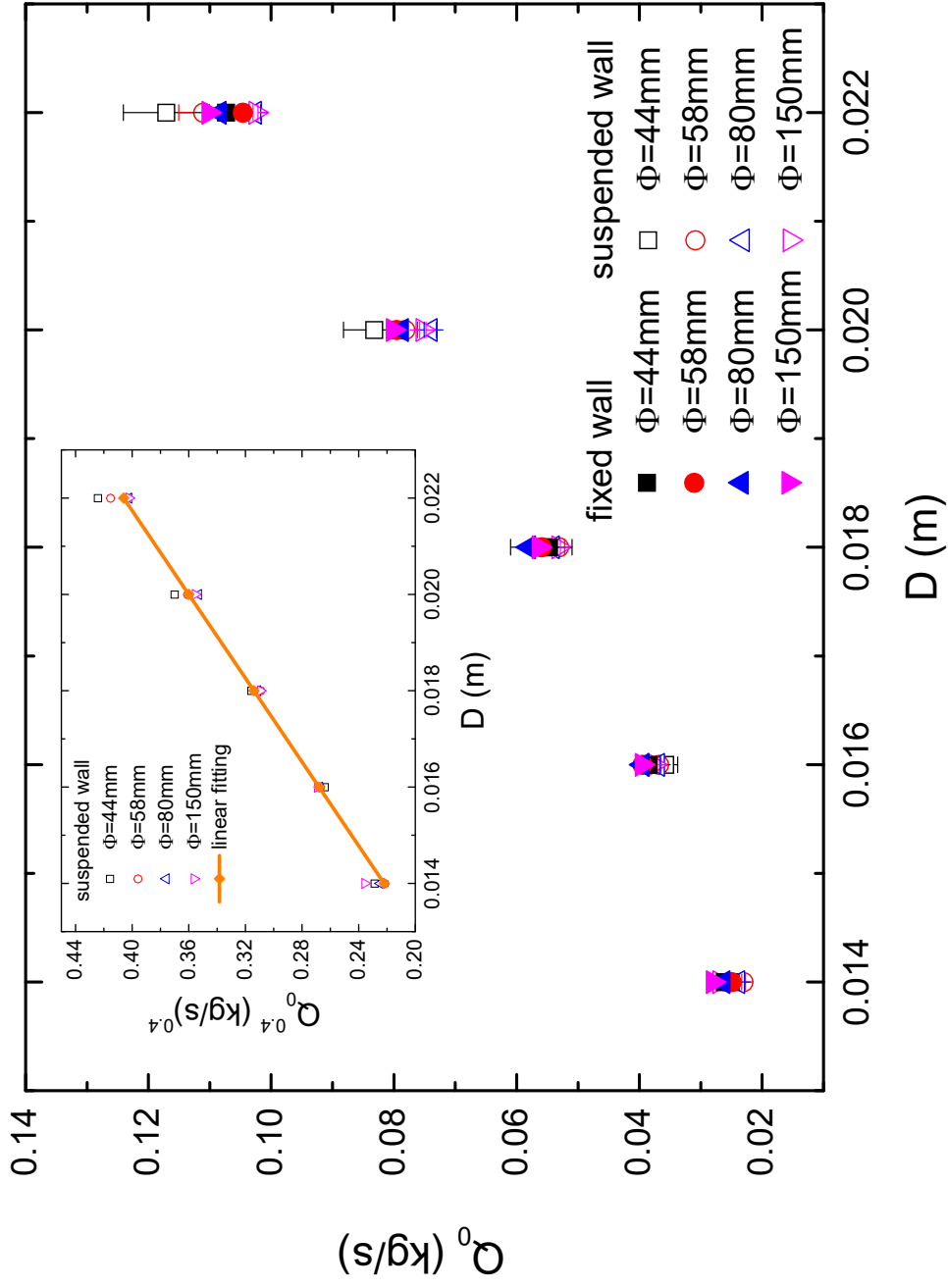

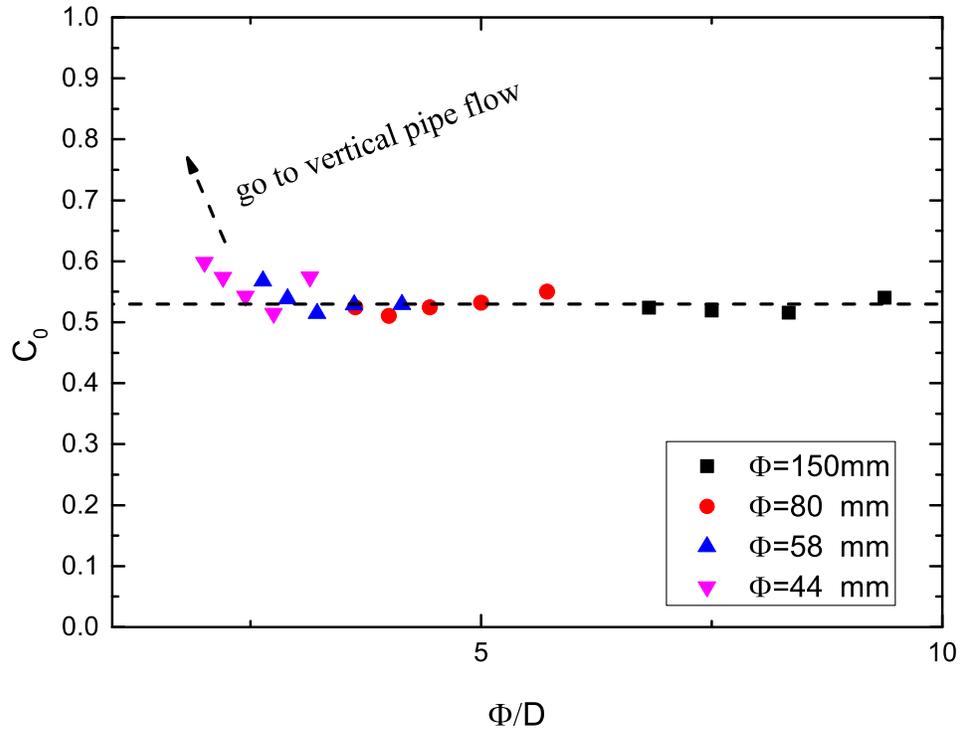